\documentstyle[twocolumn,psfig]{mn}
\newcommand{\reference}{\bibitem}
\def\beq{\begin{equation}}
\def\sigmc{\Sigma_{\rm c}}
\def\eeq{\end{equation}}
\def\bey{\begin{eqnarray}}
\def\eey{\end{eqnarray}}
\def\beqarray{\begin{eqnarray}}
\def\msun{\,{\rm {M_\odot}}}
\def\pc{\,{\rm {pc}}}
\def\eeqarray{\end{eqnarray}}

\def\sigmag{\sigma_{\rm g}}

\def\kpc{\,{\rm {kpc}}}

\def\kms{\,{\rm {km\, s^{-1}}}}
\def\vcir{V_{\rm c}}
\def\v200{V_{200}}

\def\my{{\rm M_\odot yr^{-1}}}

\title[Star formation and chemical evolution of damped Lyman $\alpha$ systems]
      {Star formation and chemical evolution of damped Lyman $\alpha$ systems}
\author[J. Ma and C. G. Shu]
{J. Ma$^{1,2,3}$, C. G. Shu$^{2,4}$\\
1. Beijing Astronomical Observatory and Beijing
Astrophysics Center (BAC) of National Astronomical Observatories,\\
Chinese Academy of Sciences, Beijing, 100012, China\\
      2. Shanghai Astronomical Observatory, Chinese Academy of
Sciences, Shanghai 200030, China\\
      3. Joint Lab of Optical Astronomy, Chinese Academy of Sciences, Beijing, 100012, China\\
      4. Max-Planck-Institute f\"ur Astrophysik
      Karl-Schwarzschild-Strasse 1, 85748 Garching, Germany\\
E-mail: majun@vega.bac.pku.edu.cn}

\date{Accepted 2000 November 8.
      Received 2000 October 10;
      in original form 2000 May 5}

\pagerange{\pageref{firstpage}--\pageref{lastpage}}
\pubyear{}

\begin{document}
\maketitle
\label{firstpage}
\begin{abstract}
In this paper, we investigate the star formation and chemical
evolution of damped Lyman-$\alpha$ systems (DLAs) based on
the disc galaxy formation model which is developed by
Mo, Mao \& White. We propose that the DLAs are
the central galaxies of less massive dark haloes present
at redshifts $z\sim 3$, and they should inhabit haloes of
moderately low circular velocity. The empirical Schmidt law
of star formation rates, and closed box model of chemical
evolution that an approximation known as instantaneous
recycling is assumed, are adopted. In our models, when the
predicted distribution of metallicity for DLAs is calculated,
two cases are considered. One is that, using the closed box
model, empirical Schmidt law and star formation time,
the distribution of metallicity can be directly
calculated. The other is that, when the simple gravitational
instability of a thin isothermal gas disc as first discussed
by Toomre is considered, the star formation occurs only
in the region where the surface density of gas satisfies the
critical value, not everywhere of a gas disc. In this case, we
first obtain the region where the star formation can occur by
assuming that the disc has a flat rotation curve and rotational
velocity is equal to the circular velocity of the surrounding
dark matter halo, and then calculate the metallicity
distribution as case one. We assume that star formation in each
DLA lasts for a period of 1 Gyr from redshifts $z=3$. There is
only one output parameter in our models, i.e. the stellar
yield, which relates to the time of star formation history and
is obtained by normalizing the predicted distribution of
metallicity to the mean value of 1/13 $Z_{\odot}$ as
presented by Pettini et al. The predicted metallicity
distribution is consistent with the current (rather limited)
observational data. A random distribution of galactic discs
is taken into account.
\end{abstract}
\begin{keywords}
galaxies: formation - galaxies: evolution -
galaxies: stellar content.
\end{keywords}
\section {Introduction}
Damped Lyman-$\alpha$ systems (DLAs) are high column density
(the HI column density is higher than $10^{20.3}$/cm$^2$) absorbers
detected in the optical spectra of quasars up to relatively
high redshift (up to $z\sim 5$), and can be used to
probe the cosmic chemical evolution, i.e. the globle evolution
of gas, metallicity and star formation rates in the Universe.
DLAs are now generally believed to be the
progenitors of present-day galactic discs (Wolfe et al. 1986; Wolfe 1988).
This view is supported by the arguments based on
mass estimates, which are from the column densities and absorber
sizes at redshifts $z\sim 2-3$, as well as by the kinematic features
consistent with models of fast-rotating (at velocities of
order 200 $\kms$), thick discs (Prochaska \& Wolfe 1997a,b, 1998;
Wolfe \& Prochaska 1998). At the same time, on the basis of
[$\alpha$/Fe] and [N/O] abundance ratios and their large scatter,
an origin of DLAs
in dwarf and low surface brightness galaxies is
also presented (Matteucci, Molaro \& Vladilo 1997;
Vladilo 1998; Jimenez, Bowen \& Matteucci 1999).
Because DLAs
contain most of the cool, neutral gas in the universe at high
redshift, they play an important role in studies of the galaxy
formation, gas process, star formation and metal enrichment as well as
the large scale structure. The pioneering work in this field was done
 by Meyer, Welty \& York (1989) and
Pettini, Boksenberg \& Hunstead (1990).
In the past few years, Pettini et al. (1994, 1997a) have devoted great
efforts in observation of DLAs' column densities and metal abundances. They
found that, with the increase of observed data,
DLAs are
generally metal poor with a mean value only of 1/13 $Z_{\odot}$ and
are very young galaxies in the early stage of chemical enrichment,
and the metallicity distribution of DLAs shows a large spread of
nearly two magnitudes. They
also pointed out that the abundance of Cr is relatively poorer than 
that of Zn
due to the depletion of dust. 

Lu et al. (1996) used the powerful Keck 10 m telescope to study the
metal abundances of many heavy
elements,
and presented a very detailed analysis of DLAs including
 comparing their distribution of
metallicities, electron density
and kinematics with those of our Milky Way galaxy,
etc.
Boiss\'{e} et al. (1998) presented HST spectra for a sample of six
DLAs with intermediate redshift ($z_{a}\le 1$), and concluded that the
available observations may be biased against dust-rich absorbers and
dust extinction causes a preferential selection of QSOs with
intervening gas relatively poor in metals, dust and molecules. They
also strongly suspect that surveys of DLAs down to fainter QSOs would
reveal more systems with large HI column densities and high
metallicities. M$\o$ller 
$\&$ Warren (1998) used HST images to
estimate the cross-section-weighted mean radius of
DLA absorbers at high redshift ($z>2$), that is
they are smaller than 23.6 $h^{-1}$ kpc
and 12.7 $h^{-1}$ kpc for $q_{0}=0.0$ and 0.5, respectively.
They also found
that for $q_{0}=0.5$ the space density of DLAs
at high redshift is more than five
times that of the spiral galaxies locally, but the problem
will be ruled out if $q_{0}$ decreases. The observations of
H$\alpha$ emission done by Bunker et al. (1999) and near-$IR$ and $IR$
done by Bechtold et al. (1998) also supported
the conclusions of M${\rm\o}$ller \& Warren
(1998) above.

Untill now, most of the researches in DLAs are for
observational constrains. There has, however, been relatively little
work on its simple prescription.
It is worthy of emphasizing that the
observation of HST (M${\rm \o}$ller \&
Warren 1998) strongly suggests DLAs
and LBGs may commonly associated with each other.
Based on their successful disc galaxy formation model,
Mo, Mao \& White (1999) have investigated
the global properties of LBGs,
such as the sizes, luminosities, kinematics and
star formation rates. 
They presented that the number of density and clustering
properties of LBGs are consistent with them
being the central galaxies of the most massive dark haloes
present at $z\sim 3$.

In this paper, under an assumption that
the DLAs are the central galaxies of less
massive dark haloes present at $z\sim 3$, and they should
inhabit haloes of moderately low circular velocity,
we study how star formation and chemical
enrichment may have proceeded in them by using
the simplest models possible with the smallest
number of parameters.

The outlines of this paper are as follows. In Section 2, we give a 
simple description of our models;
The predicted distributions of metallicity and
neutral hydrogen column
density for DLAs are
presented in Section 3,
and some correlations will also
be shown in this
section;
The summary and discussion are presented in Section 4.

\section {Modelling damped Lyman-$\alpha$ systems}

We model the assembly of DLAs in the context of
the standard hierarchical picture (White \& Rees 1978;
White \& Frenk 1991). Structure growth in this model is specified
by the parameters of the background cosmology and by
the power spectrum of initial density fluctuations.
As an illustration, we show theoretical results for a CDM model
with cosmological density parameter $\Omega_{0}=0.3$, cosmological
constant $\Omega_\Lambda=0.7$. The power spectrum is assumed to be
that given in Bardeen et al. (1986), with shape parameter
$\Gamma=0.2$ and with normalization $\sigma_{8}=1.0$. We denote
the mass fraction in baryons by $f_{\rm B}=\Omega_{\rm
B}/\Omega_0$, where $\Omega_{\rm B}$ is the cosmic baryonic
density parameter. According to the cosmic nucleosynthesis, the
currently favoured value of $\Omega_{\rm B}$ is $\Omega_{\rm
B}\sim 0.019 h^{-2}$ (Burles \& Tytler 1998), where $h$ is the
present Hubble constant in units of 100 $\rm kms^{-1}Mpc^{-1}$,
and so $f_{\rm B}\sim0.063 h^{-2}$. Whenever a numerical value of
$h$ is needed, we take $h=0.7$.

\subsection{Galaxy formation}

There are some traditional methods to investigate the galaxy formation
and evolution. One can, for instance, assume the redshift of galaxy
formation and star formation history, based on the stellar IMF and
cosmological model, to trace the observations (Tinsley 1980; Pozzetti,
Bruzual $\&$ Zamorani 1996; Jimenez, Padoan \& Matteucci
1998). White and his
collaborators 
(White $\&$ Frenk 1991; Kauffmann $\&$ White 1993; Kauffmann, White
$\&$ Guiderdoni 1993; Kauffmann et al. 1999) construct hierarchical
models of galaxy formation and evolution, and are successful in many
observations such as the statistical properties of local and distant
galaxies, etc. Madau, Pozzetti $\&$ Dickinson (1998) developed a method
which focuses on the integrated light of galaxy populations as a whole
to match the properties of the star formation history for the field
galaxies. 

In the present paper, we use the Mo, Mao $\&$ White (1998) (hereafter, MMW)
galaxy formation model in assuming that central galaxies form when collapse
of the protogalactic gas is halted either by its angular momentum or
by fragmentation as it becomes self-gravitating.
This model is in hierarchical cosmogonies,
in which a disc is a thin and centrifugally supported structure with an
exponential surface density profile. LBGs can be well understood in
MMW with large circular velocity $V_{c}$ and small spin parameter
$\lambda$ which correspond to compact objects with massive haloes (Mo et al.
1999). In this paper, we assume that DLAs
are the central galaxies of less massive dark haloes present
at redshifts $z\sim 3$, and they should
inhabit haloes of moderately low circular velocity.

According to MMW, if the mass profile of the
disc is taken to be exponential

\beq
\Sigma(r)=\Sigma_{0}\exp(-r/R_{\rm d}),
\eeq
then $\Sigma_{0}$ and  $R_{\rm d}$ that are
the central surface density and the scale length of
the exponential disc, respectively, and the rotation curve of the
galaxy, are determined uniquely.
Specifically,
\beq
R_{\rm d} \approx 8.8 h^{-1} \kpc ({{\lambda} \over {0.05}})({V_{\rm c} \over
{250\kms}}) [{H(z) \over {H_{0}}}]^{-1}
\eeq
and
\beq
\Sigma_{0} \approx 380 h \msun \pc^{-2} ({m_{\rm d} \over 0.05})({\lambda \over
0.05})^{-2} ({V_{\rm c} \over 250\kms})[{H(z) \over {H_{0}}}],
\eeq
where $m_{\rm d}$ is the fraction of disk to halo mass, $V_{\rm c}$ the
circular velocity of the halo, $\lambda$ the dimensionless spin
parameter, $H(z)$ the Hubble constant at redshift $z$, $H_{0}$ the
present-day Hubble constant (See detailed in MMW). Because the Hubble constant
$H(z)$ increases with redshift, it is
expected from equations (2) and (3) that the galaxy disc should be less
massive and smaller but have a higher surface density at higher
redshift. One can find if
$\vcir$, $\lambda$ and cosmogony are given, $R_{\rm d}$, which dominates the
disk profile, will not change.

For a given cosmogony, the mass function of dark matter haloes
at redshift $z$ can be estimated from the Press-Schechter formalism
(Press $\&$ Schechter 1974):

$$
\mbox{}\hspace{-2.0cm}{dN_{\rm h} (M_{\rm h}, z) =
-\sqrt{2 \over \pi}{\bar{\rho}_{0} \over M_{\rm h}}{\delta_{\rm c}(z) \over
\Delta(M_{\rm h})} {{\rm d}\ln\Delta(M_{\rm h}) \over {\rm d}\ln M_{\rm h}}}
$$
\beq
\mbox{}\hspace{2.2cm}{\times\exp[-{\delta_{\rm c}^{2}(z) \over
{2\Delta^{2}(M_{\rm h})}}]{dM_{\rm h} \over M_{\rm h}}},
\eeq
where $\delta_{\rm c}(z)=\delta_{\rm c}(0)(1+z)g(0)/g(z)$ is
the linear overdensity corresponding to collapse at
redshift $z$, and $g(z)$ is the
linear growth factor relative to an
Einstein-de Sitter universe
at $z$, $\delta_{\rm c}(0)\approx1.686$, $\Delta(M_{\rm h})$
is the $rms$ linear mass fluctuation at $z=0$
in a sphere which on the average contains mass $M_{\rm h}$
by $M_{\rm h}=(4\pi/3)\bar{\rho}_{0}R^{3}$, $\bar{\rho}_{0}$
is the mean mass density of
the universe at $z=0$. The relation between the mass $M_{\rm h}$ and
circular velocity $V_{\rm c}$ of a halo is in equation (2) of MMW, and
a more detailed description of the
PS formalism and
the related issues can be found in the Appendix of MMW.

  Based on the PS formalism and the $\lambda$ distribution,
with the latter being
a log-normal function with mean $\overline{\ln\lambda}=\ln0.05$ and
dispersion $\sigma_{\ln\lambda}=0.5$ (MMW), we
can generate Monte Carlo samples of the halo distribution in
the $M_{\rm h}$-$\lambda$ plane at redshifts $z=3$.
We can then use the galaxy formation model of Section 2.1 to
transform the halo population into a DLA population.
Finally, suppose that each halo at $z\sim 3$ has a central galaxy with
a star formation rate (SFR) that is a monotonic function
of halo mass, and only a negligible fraction of haloes
hosts a second galaxy bright enough to be seen.
The observed DLAs then correspond to haloes at $z\sim 3$.
Here a lower-limit for $V_{\rm c}$ of 50 km/s of haloes is chosen,
and a random
distribution of galactic discs is taken into account.

It should be pointed out that the interaction between discs and bulges is
not considered in the present study. This interaction will become more
and more important for very 
compact objects which correspond to $\lambda<0.025$ (MMW). Here, we
treat galaxies with $\lambda<0.025$ as  
$\lambda=0.025$. It can be easily estimated from Warren et
al. (1992), Cole $\&$ Lacey (1996) and Steinmetz $\&$ Bartelmann
(1995) that,  there are only 10 percent galaxies with $\lambda<0.025$.
This will not influence our main results.

\subsection{Star formation rates}

For star formation rates (SFR), we take the empirical Schmidt law (Kennicutt
1998)
\beq
\Sigma_{\rm SFR} = a ({{\Sigma_{\rm g} (r)} \over
{1 \msun {\rm pc}^{-2}}})^{b} \my{\rm pc}^{-2} ,
\eeq
where
\beq
a = 2.5 \times 10^{-10}, b=1.4,
\eeq
respectively. Here $\Sigma_{\rm SFR}$ is the SFR per unit area, and
$\Sigma_{\rm g} (r)$ the gas surface density.
This star formation law was derived by averaging the
cool gas density over large areas on spiral
discs and over starburst regions.

\subsection{Closed box model}

For a closed box model, the metallicity $Z$ of gas is
(Binney \& Merrifield 1997; Tinsley 1980)
\beq
Z-Z_{i} = -y\ln(\Sigma_{\rm g} (r)/\Sigma_{\rm g0} (r)) = y\ln\mu^{-1},
\eeq
where $Z_{i}$ is the initial  metallicity of cool gas that
is taken to be 0.01$Z_{\odot}$ in the present paper, $y$ the stellar
yield that is obtained by normalizing the
predicted distribution of metallicity to
the mean value of 1/13 $Z_{\odot}$ as presented by Pettini et al. (1997a)
in this study
(see next section),
$\Sigma_{\rm g0}(r)$ the initial gas surface density and $\mu$ the gas
fraction, and
\beq
\frac{{\rm d}\Sigma_{\rm g} (r)}{{\rm d}t}=-(1-R_{\rm r})\Sigma_{\rm SFR},
\eeq
where $R_{\rm r}$ is the returned fraction of stellar mass into the ISM,
and we take $R_{\rm r}=0.3$ for a Salpeter IMF (see Madau et al. 1998).
We assume that at initial time ($t_\star=0$), $\Sigma_{\rm SFR}=0$ and
$\Sigma_{\rm g0} (r)=\Sigma_0 (r)\exp(-r/R_{\rm d})$,
with the latter being given by equation (1).
After a period of $T$, the surface density of the survived
cool gas can be obtained by equations (5) and (8), that is
\beq
\Sigma_{\rm g} (r)= \Sigma_{\rm g0} (r) (1+T/\tau)^{-{1 \over {b-1}}},
\eeq
and the current SFR is
\beq
\Sigma_{\rm SFR} = a {\Sigma _{\rm g0}^b (r)} (1+T/\tau)^{{{b} \over {1-b}}} ,
\eeq
where
\beq
\tau = {\Sigma_{g0}^{1-b} (r) \over {a (b-1)(1-R_{\rm r})}}.
\eeq

As we know,
in the simple closed-box model of chemical evolution, all metals produced
by stars are retained within the galaxy. The metallicity of the galaxy
is simply determined by the stellar yield $y$, defined as the mass of
metals produced per solar mass of long-lived stars that are formed
(Tinsley 1980).
It is worthy to note that, the instantaneous recycling approximation
is not valid as regards the contributions to metal
enrichment of either type Ia supernovae or
of stars of a few solar masses (Binney \& Merrifield, 1997). 
Pagel (1997) has refined to take into account the time delay between
star formation and the ejection into the ISM of heavy
elements.

In this paper, we use [X/H]\footnote{We use
the normal notation where $[X/Y]=\log (X/Y)-\log (X/Y)_{\odot}$.}
for the logarithm metallicity,
and all the histograms are normalized to unity.

\subsection{Criterion for star formation}

According to Toomre (1964), in the case of a gaseous disc,
instability for star formation is expected if the surface density of gas exceeds
the following critical value

\beq
\sigmc (r)=\frac{\alpha\kappa\sigmag}{3.36G},
\eeq
where $\alpha$ is taken to be 2
(van der Kruit \& de Grijs, 1999); and
\beq
\kappa=1.41\frac{V}{r}{(1+\frac{r}{V}\frac{{\rm d}V}{{\rm d}r})}^{1/2};
\eeq
and $\sigmag$ is a gas velocity dispersion, we take it to be
$6\kms$ (Kauffmann 1996),
which is consistent with the observational
results at any redshift till now (see Larson 1998).
 
When this criterion is considered, the star formation occurs only in the
region where the surface density of gas satisfies the critical value,
not everywhere of the gas disc.

\section{Results}

In this section, we present the predicted normalized distributions
of metallicity and HI column density for the DLA population.
As explained
by Pettini, King \& Smith (1997b), [Zn/H] is a straight forward
measure of the degree of metal enrichment analogous
to the stellar [Fe/H]. So, we use the mean value of [Zn/H]
as presented by Pettini et al. (1997a) to normalize
the predicted distribution of metallicity for DLAs
in order to obtain the steller yield. A random
distribution of discs is taken into account.

\subsection{Distribution of metallicity without considering
the Toomre's criterion}

At first, we do not consider the criterion for star formation
which is developed by Toomre (1964). Based on the Monte Carlo samples
at $z=3$, we can calculate the metallicity $Z$ of DLA population by the
formulas from (5) to (11) by assuming that,
at initial time ($t_\star=0$), $\Sigma_{\rm SFR}=0$ and
$\Sigma_{\rm g0} (r) = \Sigma_0\exp(-r/R_{\rm d})$;
and the star formation time $T$ is 1 Gyr.
$\Sigma_0$ and $R_{\rm d}$ are decided by
equations (2) and (3).
The stellar yield is obtained by normalizing the
predicted distribution of metallicity to
the mean value of 1/13 $Z_{\odot}$ as presented by Pettini (1997a).
It is 0.22 $Z_\odot$.
The results are plotted in Fig. 1. For comparison,
the metallicities of 18 DLAs at redshifts $z \sim 3$ $ (2.3<z<3.4)$
that were obtained by
Meyer, Welty \& York (1989) and
Pettini et al. (1994, 1997a)
\footnote{The solar
system abundance of Zn is from the
compilation by Anders \& Grevesse (1989),
i.e. $\log (\rm{Zn/H})_{\odot}=-7.35$.}
are also presented. This figure shows that,
the observational data that have
a spread of nearly two orders of magnitude
agree with our model, but the predicted
metallicity distribution cannot
spread as widely as the observations.
The observational data are still quite sparse,
and larger and more complete samples are needed.
In order to test whether our model fits the abundances of
DLAs, we provide the Kolmogorov-Smirnov Test.
The results are that the maximum value of the absolute difference
is 0.33, and the significance level probability is 69.94 percent.

\begin{figure}
\centerline{\psfig{figure=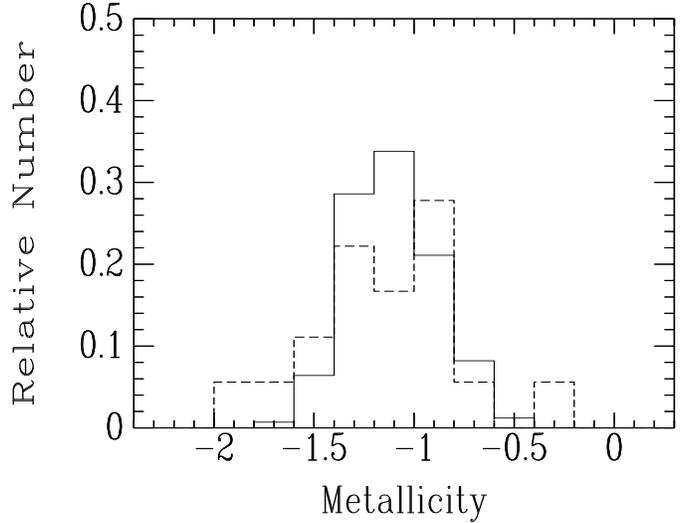,angle=-90,height=8.0cm}}
\caption{The distribution of metallicities for DLAs. The
solid histogram gives the model prediction and the
dashed histogram shows the observational data in Meyer et al. (1989)
and Pettini et al. (1994, 1997a).}
\end{figure}

\subsection{Distribution of metallicity with considering
the Toomre's criterion}
 
When we take into account the Toomre instability criterion
of star formation, the star formation occurs only in the region
where the surface density of gas must be larger than $\Sigma_c (r)$.
So, we first obtain the regions where the star formation
can occur based on the formulas (1), (12) and (13), assuming
that the disc has a flat rotation curve and
rotational velocity is equal to the circular velocity
of the surrounding dark matter halo.
If the observation is not in these regions,
the predicted distribution of metallicity for DLA population
is initial one of the cool gas,
i.e. 0.01 $Z_{\odot}$.
When the observation is just in these regions,
the metallicity of the DLA population can be obtained as section 3.1.
In this case, the stellar yield is 0.70 $Z_\odot$.
Fig. 2 presents the normalized model distribution of metallicity.
For comparison,
the metallicities of 18 DLAs at redshifts $z \sim 3$ $ (2.3<z<3.4)$
that were obtained by
Meyer, Welty \& York (1989) and
Pettini et al. (1994, 1997a) are also plotted.
This figure shows that,
the model data which spreads more than two magnitudes can
match the observed ones (Meyer et al. 1989, Pettini et al. 1997a),
although the model is a little metal rich.
As in Fig. 1, we also
provide the Kolmogorov-Smirnov Test.
The results are that the maximum value of the absolute difference
is 0.25, and the significance level probability is 84.75 percent.

\begin{figure}
\centerline{\psfig{figure=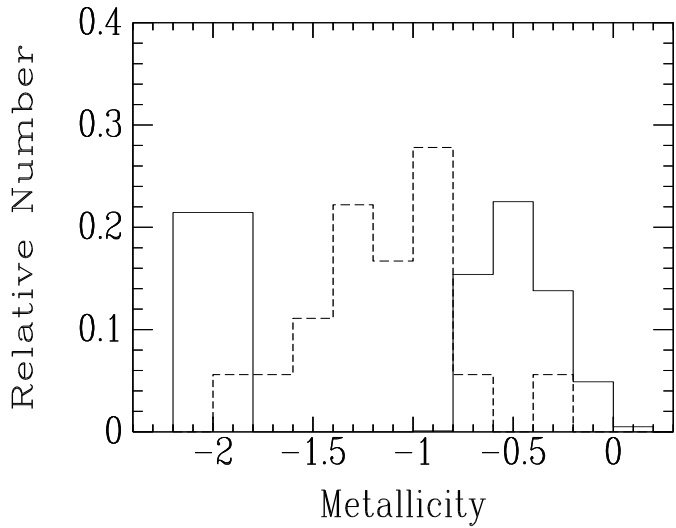,angle=-90,height=8.0cm}}
\caption{The distribution of metallicities for DLAs. The
solid histogram gives the model prediction and the
dashed histogram shows the observational data in Meyer et al. (1989)
and Pettini et al. (1994, 1997a).}
\end{figure}

\subsection{Distribution of HI column density}

HI column density is the major characteristics of DLAs, because it
is the first physical parameter that we can obtain from observations of QSO
absorption lines. If we assume that the survived
gas from the star formation in a disc is dominated by the
neutral hydrogen, the model
column density of neutral hydrogen can be obtained
by the formulas from (5) to (11).
The predicted distribution of HI column density for DLAs
without and with considering the Toomre's criterion are
shown in Figs. 3 and 4, respectively. For comparison,
the HI column densities of 18 DLAs at redshifts $z \sim 3$ $ (2.3<z<3.4)$
that were obtained by
Meyer, Welty \& York (1989) and
Pettini et al. (1994, 1997a) are also plotted.
These two figures present that,
the observational neutral hydrogen column densities spread
narrower than the predicted ones. The reason is,
perhaps, that the extinction may results in a preferential
selection of QSOs with systems displaying a low HI and H$_2$ content.

In fact, Boiss\'{e} (1995) has already noticed that,
the absence of very large N(HI) values (e.g. $\geq 10^{22}$ cm$^{-2}$)
which would be quite easily recognized in the numerous
low resolution optical spectra obtained so far,
supports the reality of cut-off caused by extinction.

Surveys of DLAs samples drawn from the observation of
fainter QSOs may reveal more systems with large N(HI), and
spectroscopic surveys of fainter QSOs can lead to a
more representative view of the whole diversity of the DLAs.

\begin{figure}
\centerline{\psfig{figure=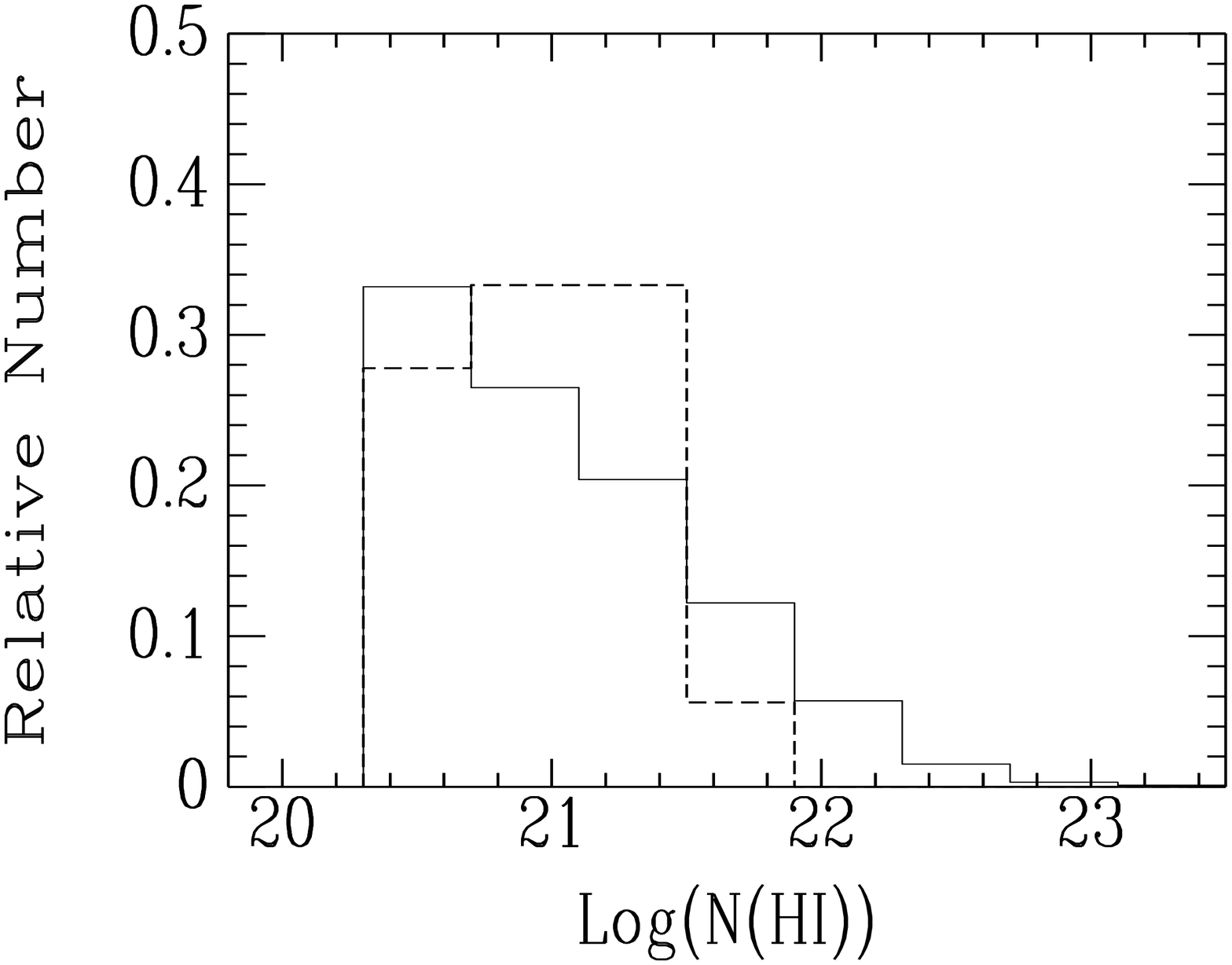,angle=-90,height=8.0cm}}
\caption{The distribution of HI column densities for DLAs. The
solid histogram gives the model prediction and the
dashed histogram shows the observational data in Meyer et al. (1989)
and Pettini et al. (1994, 1997a).}
\end{figure}

\begin{figure}
\centerline{\psfig{figure=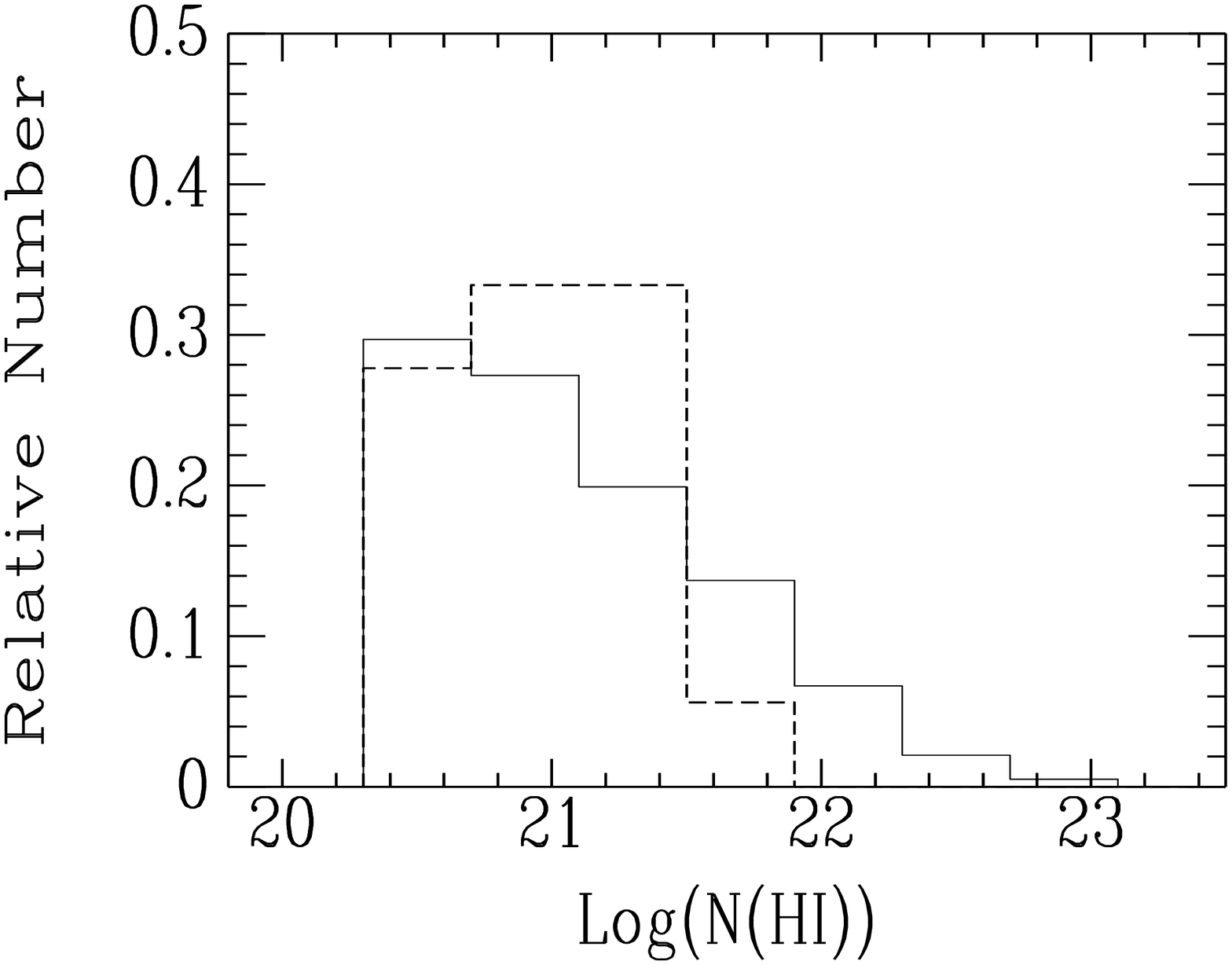,angle=-90,height=8.0cm}}
\caption{The distribution of HI column densities for DLAs. The
solid histogram gives the model prediction and the
dashed histogram shows the observational data in Meyer et al. (1989)
and Pettini et al. (1994, 1997a).}
\end{figure}

\subsection{Relation between metallicity and HI column density}

In order to present the predicted relation between metallicity and HI
column density for DLAs, we randomly select 100 samples from
the DLA population  by their cross-sections
because the observations of DLAs are based on the QSO absorption lines.

\begin{figure}
\centerline{\psfig{figure=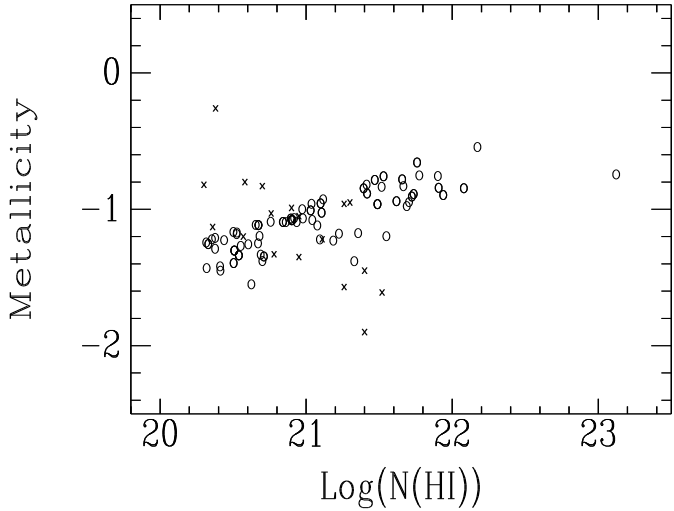,angle=-90,height=8.0cm}}
\caption{The relation between metallicity and
HI column density for DLA population.
The blank circles and crosses represent the model prediction
and the observational data in Meyer et al. (1989)
and Pettini et al. (1994, 1997a), respectively.}
\end{figure}

\begin{figure}
\centerline{\psfig{figure=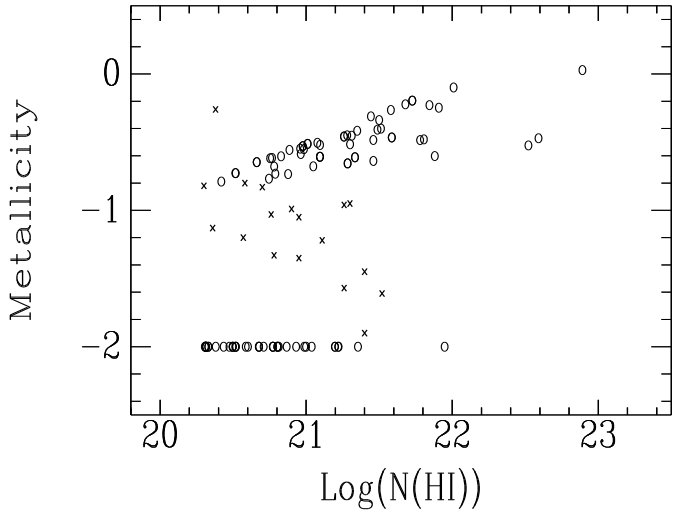,angle=-90,height=8.0cm}}
\caption{The relation between the metallicity and
HI column density, where
blank circles and crosses represent the model prediction
the observational data in
Meyer, Welty \& York (1989) and Pettini et al.
(1994, 1997a), respectively.}
\end{figure}

In Figs. 5 and 6, we plot the predicted distribution of
[Zn/H] as a function of HI column density
for DLAs without and with considering the Toomre's criterion,
respectively. For comparison,
the observational data of 18 DLAs at redshifts $z \sim 3$ $ (2.3<z<3.4)$
that were obtained by
Meyer, Welty \& York (1989) and
Pettini et al. (1994, 1997a) are also plotted.
The predicted correlations (especially,
without considering the Toomre's criterion)
are that, the
higher the column density, richer the metallicity.
In fact, Boiss\'{e}
et al. (1998)
have already emphasized that, in an unbiased sample, the large N(HI) values
should correspond to the innermost parts of galaxies
where the metallicity is presumable higher.
The observational data show a different trend, i.e.
the metallicity is
richer with the decreasing of column density because of
the extinction bias as pointed out by Boiss\'{e} et al. (1998).

\section{Summary and Discussion}

In the present paper, we establish two simple scenarios to investigate
the nature of DLAs at redshifts $z\sim3$ based on the disc
galaxy formation model developed by Mo, Mao \& White (1998). One is
without considering the Toomre's criterion,
another is with considering the Toomre's criterion. The star formation rate is
chosen to be in the form of the empirical Schmidt law (Kennicutt
1998). The approximation of
instantaneous recycling is assumed.
The predicted distribution of metallicity for DLAs spreads
larger than one order of magnitude, which is consistent with the
observations by Meyer, Welty \& York (1989) and
Pettini et al. (1994, 1997a).

Although the star formation time of 1 Gyr, which
we adopted, is arbitrary, its
order is consistent with that of observations (Bechtold et al. 1998). It is
no problem for us to choose it as a typical time of star formation to
investigate the global properties of DLAs.

In our models, the star formation time scale in the disc is
implicitly assumed to be the same everywhere. This will lead to the
slow consumption of gas in the central parts of the disc.
In fact, as we know well, the star formation in the central
region is very rapid, and the central region should correspond
to higher metallicities and
lower HI column densities.
Furthermore, we treat all the
survived gases as HI. There actually 
exist ${\rm H}_{2}$, ionized gases and other elements. This is also leads to
a overestimation of HI column density, especially in the central
region of the disc. 
In addition, the DLAs that correspond to have small $V_{\rm c}$,
should form earlier and the merging
could take place more frequently.
So, the DLAs are to a certain extent more
complicated.

  There is only one output parameter in our models, i.e. the
stellar yield, which relates to the star formation time scale and
is obtained from the normalization of the predicted distribution of
metallicity
to the mean value of 1/13 $Z_{\odot}$ as presented by Pettini et al. (1997a).
The metallicity of the initial
cool gas is assumed to be of 0.01 $Z_{\odot}$.
According to Pagel (1987), the stellar yield
is 0.3 $Z_{\odot}$ for the disc clusters and 0.4 $Z_{\odot}$ for the solar
neighborhood.
The stellar yields in our models are also quite similar to that of LMC of some
0.2 $Z_{\odot}$ (Binney $\&$
Tremaine 1987).

Finally, we must point out, although our models are very simple without many
physical processes being considered, such as the interaction between
discs and bulges, merging effects, gas reheating and supernovae
feedback, etc., it can predict some properties of DLAs.

\section*{Acknowledgments}

We are indebted to the anonymous referee for
many critical comments and helpful suggestions that
have greatly improved our paper.
We are also grateful to
H. J. Mo for much help in finishing this paper.
This project is partly supported by Chinese National Natural Foundation and 
the ``Sonderforschungsbereich 375-95 f\"ur Astro-Teilchenphysik'' der
Deutschen Forschungsgemeinschaft. 

\vfill\eject
{}
\end{document}